\documentclass[prb,twocolumn,showpacs,superscriptaddress,preprintnumbers,amssymb]{revtex4-2}
\usepackage{graphicx}
\usepackage{float}
\usepackage{color}
\usepackage{dcolumn}
\usepackage{bm}

\newcommand{\beq}{\begin{equation}}
\newcommand{\eeq}{\end{equation}}
\newcommand{\beqn}{\begin{eqnarray}}
\newcommand{\eeqn}{\end{eqnarray}}

\newcommand{\ra}{\rightarrow}

\newcommand{\cA}{ {\cal A} }

\newcommand{\cP}{ {\cal P} }
\newcommand{\cS}{ {\cal S} }
\newcommand{\cV}{ {\cal V} }

\newcommand{\cQ}{ {\cal Q} }

\newcommand{\vect}[1]{{\bm{#1}}}

\newcommand{\ii}{\mathrm{i}}

\newcommand{\SU}{\mathrm{SU}}
\newcommand{\U}{\mathrm{U}}

\begin{document}

\title{``Symmetry-from-Anomaly" in Condensed Matter related Constructions}

\author{Cenke Xu}
\affiliation{Department of Physics, University of California, Santa Barbara, CA 93106}

\date{\today}

\begin{abstract}

The noninvertible axial symmetry constructed from the ABJ-anomaly has attracted enormous interest. We discuss the mechanism of ``symmetry-from-anomaly" in condensed matter-related models in both $1d$ and $3d$ spaces (which correspond to $(1+1)d$ and $(3+1)d$ space-time). {\it Within the models discussed here}, we establish the connection between field theory quantities such as different versions of the axial charge, and quantities with simple physical meanings in our systems. In our models and likely a class of related constructions, the existence of a topological order is necessary for the purpose of properly defining the axial symmetry. But the proper axial symmetry we define, though requires a topological order, is different from the noninvertible axial symmetry discussed in recent proposals.

\end{abstract}

\maketitle

\section{Introduction}

Let us review the basic logic of the relation between the ABJ-anomaly and noninvertible symmetry. We consider the $(3+1)d$ massless QED with one Dirac fermion. The axial charge of the Dirac fermion is not conserved when it is coupled to a U(1) gauge field, due to the ABJ-anomaly~\cite{adler1,adler2}: \beqn \partial_\mu j^{A,\mu} = - \frac{e^2}{16\pi^2} \epsilon^{\mu\nu\rho\sigma} F_{\mu\nu} F_{\rho\sigma} = - \frac{e^2 }{2\pi^2} \vect{E} \cdot \vect{B}. \eeqn Though this anomaly does not cause any inconsistency of the theory, it does explicitly break the global axial symmetry. But one can define another modified axial charge, which is formally conserved~\cite{adler1}: \beqn \tilde{Q}^A = Q^A + \int d^3x \ \frac{e^2}{4\pi^2} \epsilon_{ijk}A_i \partial_j A_k. \eeqn However, problem emerges when we perform axial rotation with the conserved charge $\tilde{Q}^A$: the angle $\alpha = 2\pi /N$ axial rotation defined with the new axial charge $ \exp(- \frac{\ii }{2} \frac{2\pi}{N} \tilde{Q}^A)$ is not gauge invariant. In particular, if we perform the angle $\alpha = 2\pi /N$ axial rotation to a subregion $\cV$ in the $3d$ space, at the boundary of the subregion the system acquires an extra level-$1/N$ Chern-Simons term, which is not properly quantized: \beqn \tilde{S}_{\partial \cV} = \int_{\partial \cV} d^2x dt \ \frac{1}{N} \frac{e^2}{4\pi} \epsilon^{\mu\nu\rho}A_\mu \partial_\nu A_\rho . \label{axialcs0}\eeqn Therefore the fractional axial charge $\pi Q^A/N$ is {\it gauge invariant but not conserved}; while the modified fractional axial charge $\pi \tilde{Q}^A/N$ is {\it conserved but not gauge invariant}.

It was proposed that there exists a proper axial $\cQ^{\cA}$ which is conserved, and also generate a gauge invariant $\alpha = 2\pi /N$ axial rotation. But the price we pay is that, the proper axial rotation must be noninvertible~\cite{shao2022,cordova2022}, as it includes a level-$N$ Chern-Simons term (hence a noninvertible fractional quantum Hall state) on the $(2+1)d$ boundary of the subregion $\cV$: \beqn \cS_{\partial \cV} = \int_{\partial \cV} d^2x dt \ \frac{e }{ 2\pi} \epsilon^{\mu\nu\rho} a_\mu \partial_\nu A_\rho - \frac{N }{4\pi} \epsilon^{\mu\nu\rho}a_\mu \partial_\nu a_\rho. \label{axialcs}\eeqn
The connection between the ABJ-anomaly and noninvertible symmetry is intriguing. But the explicit physical mechanism for the decoration of each domain wall of the axial rotation with a fractional quantum Hall state needs further efforts.

In this work we consider condensed matter motivated models, and we would like to realize the $Z_N$ axial rotation as a {\it natural} symmetry of our systems. In particular, we discuss a $1d$ conductor, and a $3d$ Weyl semimetal (which are in $(1+1)d$ and $(3+1)d$ spacetime)~\cite{weyl1,weyl2}, both are coupled to the electromagnetic (EM) field. Most of the time we will consider the simplest scenario where the ABJ-anomaly plays a nontrivial role, and take the standard approximation in solid state physics: the EM field is treated as an external field with low strength, slow dynamics, and long wavelength modulation. Due to the 't Hooft anomaly of the axial symmetry, it cannot be realized as an onsite $\U(1)$ symmetry in principle~\cite{nn1,nn2}. In our models the axial symmetry is implemented as translation symmetry, which is the most natural realization of axial symmetry in condensed matter systems, such as the Weyl fermions in the Weyl-semimetal~\cite{weyl1,weyl2} and $^3$He-A superfluid~\cite{he3}. The axial symmetry charge is therefore realized as the mechanical momentum of the charge carriers.

We note that for a free fermion system, infinite numbers of non-onsite charges with the form $\sum_k f(k) c^\dagger_k c_k$ are conserved. But with generic interactions, if one only assumes lattice translation symmetry, the crystal momentum is the only quantity that is generically conserved without fine-tuning, though it is conserved mod the reciprocal momentum.

The ABJ-anomaly and the Schwinger anomaly exist in our models, hence the axial charge $Q^A$ is not conserved. We will propose our way of defining a proper axial charge $\cQ^\cA$ that generates a gauge invariant $\alpha = 2\pi /N$ axial rotation while being conserved by itself. Some ``dark sector" $\psi_{\rm d}$ needs be introduced in our system, in addition to the gapless QED. The dark sector needs to meet the following criteria:

\begin{itemize}

\item It is ``dark" in the sense that it does not affect the low energy physics, i.e. the low energy physics is still captured by the $(3+1)d$ QED with one Dirac fermion. This means that the dark sector is gapped, and does not break the key symmetry that keeps the QED gapless.

\item It contributes to the total axial charge, in particular, it should formally compensate the domain wall $\partial \cV$ with a Chern-Simons term in Eq.~\ref{axialcs0}, in a gauge invariant way.

\item The axial rotation defined with the proper axial charge of the entire ``electron+$\psi_{\rm d}$" system should keep the bulk of $\cV$ unchanged, as the proper axial rotation is a real symmetry.

\end{itemize}

In order to properly define the axial charge, a noninvertible topological order will be introduced in our constructions. However, in our models the proper axial rotation generated by $\cQ^\cA$ should {\it not} be viewed as a noninvertible symmetry, as the topological order resides in the $(3+1)d$ spacetime bulk rather than the $(2+1)d$ domain wall $\partial \cV$.

\section{Basic notions}

Let us first perform a simple practice,
and consider a $1d$ system with $N_e$ electrons in an electric field. The Hamiltonian in the first-quantized form is \beqn H = \sum_{i = 1}^{N_e} \frac{p_{e,i}^2}{2m_e} + e V(x_{e,i}). \eeqn We consider a uniform electric field, i.e. $V(x_{e,i})$ is a linear function $\sum_{i} - E x_{e,i} $. The Hamiltonian is clearly not invariant under translation $x_{e,i} \ra x_{e,i} + b$, and it acquires an extra term: \beqn \delta H = \sum_{i = 1}^{N_e} V(x_{e,i} + b) - V(x_{e,i}) = - (e N_e b) E. \eeqn $eN_eb$ is the total electric charge polarization generated by translation, as the polarization is defined as $\int dx \rho_e(x) x$.

The correction $\delta H$ is in fact a topological $\Theta$-term of the $(1+1)d$ EM field: \beqn \delta \cS = \int dx dt \ \frac{\Theta}{2\pi} \epsilon^{\mu\nu} F_{\mu\nu} = \int dt \int dx \frac{\Theta}{\pi} E, \eeqn where $\int dx \frac{\Theta}{\pi}$ is the total polarization. Even in this simple practice problem we can see the analogy with the ABJ-anomaly in $(3+1)d$: the translation symmetry is broken by a uniform electric field, and it generates a topological $\Theta$-term, just like the axial rotation of the $(3+1)d$ QED. The physical meaning of a $(1+1)d$ $\Theta$-term is just the net charge polarization.

Now we add $N_h$ holes to the system: \beqn H = \sum_{i = 1}^{N_e} \frac{p_{e,i}^2}{2m_e} + e V(x_{e,i}) + \sum_{i = 1}^{N_h} \frac{p_{h,i}^2}{2m_h} - e V(x_{h,i}). \eeqn
Translation of both electrons and holes can still keep the total Hamiltonian invariant, but this only happens when
$N_e = N_h$, i.e. the entire system is charge neutral. Hence in this case the translation is still a symmetry,
and the total momentum of the entire neutral system should be a conserved quantity in a uniform electric field. Also, if we translate both electrons and holes, there is no total polarization generated, i.e. the topological $\Theta$-term generated by electrons and holes are canceled out. These results hold even with interactions.

$ $

Now we can review some basic notions of single-particle quantum mechanics coupled with the EM field. In quantum mechanics, the mechanical momentum of a charge $q$ particle is defined as \beqn \vect{P} = \vect{p} - q \vect{A}(\vect{x}), \eeqn where $\vect{p}$ is the ordinary canonical momentum $\vect{p} = - \ii \vect{\nabla}$. Unlike the canonical momentum, the mechanical momentum is fully gauge invariant, and it generates a spatial translation attached with a Wilson-line, i.e. the ``parallel transport" on a $\U(1)$ bundle.

The mechanical momentum $\vect{P}$ is not conserved in a nonzero magnetic and electric field $\vect{E}$. In the temporal gauge $A_0 = 0$, the evolution of $\vect{P}$ in a nonzero electric field comes from the Maxwell equation of $\vect{A}$ in the expression of $\vect{P}$: \beqn \frac{d\vect{P}}{dt} = \ii [H, \vect{P}] + \frac{\partial\vect{P}}{\partial t} = - q \frac{\partial \vect{A}}{\partial t} = q \vect{E}. \eeqn Note that with a $\vect{E}$ field, the mechanical momentum $\vect{P}$ is {\it gauge invariant but not conserved}.

With a {\it uniform} electric field, the total mechanical momentum $\vect{P}^T$ of an entire charge neutral system should be both gauge invariant and also conserved, such as the situation of a compensated band structure when electrons and holes have equal density. Generally speaking, in a system with translation symmetry, the total mechanical momentum (rather than momentum of a subsystem) should be the conserved quantity associated with translation. This is still true with interactions.

In our construction we treat electrons (coupled with the EM field) as our target system whose low energy physics is the QED, but we will also introduce a dark sector. The dark fermion $\psi_{\rm d}$ always carries the opposite charge from the electron. Within the models discussed in this work, we make the following identification :

\begin{itemize}

\item The original fractional axial charge $\pi Q^A/N $ corresponds to the mechanical momentum of gapless electrons, which is gauge invariant but not conserved.

\item
The proper fractional axial charge $\pi \cQ^{\cA} / N $ corresponds to the total mechanical momentum of the electron+$\psi_{\rm d}$ system, and it is both conserved and gauge invariant.

\end{itemize}

\section{A 1d model}

\subsection{Axial charge $Q^A$}

We first consider a $1d$ metal with electrons and dark fermion $\psi_{\rm d}$ coupled with the EM field, in this example $\psi_{\rm d}$ can be viewed as the hole. We assume the entire system is ``compensated", so the total density of electric charges is zero. The lattice version of the model will be discussed in the appendix. Here we assume that at low filling factor, all the relevant modes have long wavelengths, which facilitates a description in the continuum. We first discuss physics of electrons, which will be our target system: \beqn H &=& \int dx \ - \frac{1}{2m} \psi^\dagger \left( \nabla_x - \ii e A(x) \right)^2 \psi - \mu \psi^\dagger \psi. \label{H} \eeqn This continuum Hamiltonian is meant to be a description at the intermediate scale. The quantity that we will focus on is the mechanical momentum of the electrons: \beqn P = \int dx \ \cP(x) = \int dx \ \psi^\dagger (- \ii \nabla_x - e A(x)) \psi.\eeqn We can expand $P$ with the low energy modes, i.e. expand at the two Fermi wave vectors $\pm k_F$ of $\psi$: \beqn \cP(x) &\sim& k_F (\psi_R^\dagger \psi_R - \psi^\dagger_L \psi_L) - e A(x) \psi^\dagger \psi \cr\cr &=& k_F (\rho_R - \rho_L) - e A (x) \rho. \label{momentum2}\eeqn Let us also assume that at low energy the fermion density is uniform with filling factor $\nu = 1/N$, and the relation between $k_F$ and density is determined by the Luttinger theorem: $k_F = \pi \nu = \pi/N$. Therefore the mechanical momentum further reduces to \beqn \cP &=& \frac{\pi}{N} \rho^A, \cr\cr \rho^A &=& (\rho_R - \rho_L) - \frac{e}{\pi}A(x). \label{axial}\eeqn As we discussed before, the mechanical momentum is gauge invariant. We will identify the operator $Q^A = \int dx \rho^A$ in Eq.~\ref{axial} as the axial charge, i.e. the mechanical momentum $P$ provides a gauge invariant regularization of the axial charge $Q^A$.

We need to derive the time-evolution of the mechanical momentum $P$, and demonstrate that it is indeed the axial charge $Q^A$, whose evolution is governed by the Schwinger-anomaly equation in the field theory. The derivation is straightforward in the continuum model, and a derivation from the lattice model is in the appendix. The bottom line is that, the gauge invariant $Q^A(x)$ defined above is not conserved, as it evolves under Eq.~\ref{H} with nonzero coupling constant $e$: \beqn \frac{dP}{dt} = \ii [H, P] - \int dx \ e \frac{\partial A}{\partial t}\rho(x) = \int dx \ e E(x) \rho(x). \eeqn From the evolution of $P$, we obtain the evolution of axial charge $Q^A = \int dx \ \rho^A(x)$: \beqn \frac{dQ^A}{dt} = \int dx \ \frac{e}{\pi} E(x), \eeqn which has the same expression as the Schwinger-anomaly, and it is essentially the Newton's law. Again, we have used the fact that $\nu = 1/N$ and $k_F = \pi/N$ which is fixed by the Luttinger theorem.

Note that just like the mechanical momentum, the definition of $Q^A$ already involves the gauge field $A(x)$. This is consistent with the Abelian bosonization, where the gauge field is included in $\rho^A(x)$ explicitly: \beqn \rho^A(x) \sim \frac{1}{\pi} \left( \nabla_x\theta - e A(x) \right). \eeqn In fact, in field theory, in order to define a gauge invariant $\rho^A(x)$, we also need to insert a short Wilson-line after point-splitting.

The mechanical momentum is the generator for parallel transport, which is the gauge invariant translation in this system. Now let us investigate the partial translation operator on subregion $(-\infty, x)$: \beqn && \exp \left( \ii \int_{-\infty}^x dx \ \cP(x) \right) \cr\cr &=& \exp \left( \ii \int_{-\infty}^x dx \ \frac{1}{2} \frac{2\pi}{N} \rho^A(x)\right) \cr\cr &\sim& \exp \left( \ii \frac{1}{N} \theta(x) - \ii \int_{-\infty}^x dx \frac{ e}{N} A(x) \right). \label{fraction}\eeqn The partial translation operator becomes the axial rotation at angle $\alpha = 2\pi/N$, and it creates a fractional charge $e/N$ located at position $x$, attached with a fractional Wilson line. The entire operator is still gauge invariant.

The existence of fractional charges discussed in this section may seem puzzling. In fact, Eq.~\ref{fraction} is just a translation on a subregion $(-\infty, x)$. At fractional filling $\nu = 1/N$ of a $1d$ system, it is well-known that translation on a subregion creates fractional charge $1/N$~\cite{fraccharge1,fraccharge2,fraccharge3}. However, this is not a deconfined mobile fractional charge, it is a ``polarization charge", meaning this charge is at the end of a string of charge-dipoles. Generally speaking, a polarization charge can take any value, and it does not require a deconfined topological order. Later we will discuss the condition when a topological order is needed.

We note that the modern quantum theory of polarization is formulated
based on the Berry phase of the Bloch states~\cite{Polaberry0,Polaberry1,Polaberry2,Polaberry3,Polaberry4,Polaoshikawa}, at least for noninteracting fermions. The charge polarization is proportional to the $\Theta$-term of the EM field in $(1+1)d$.

\subsection{Modified axial charge $\tilde{Q}^A$ after restoring conservation}

Given the Schwinger-anomaly, one can define a modified axial density $\tilde{Q}^A$: \beqn \tilde{Q}^A = Q^A + \int dx \ \frac{e}{\pi} A(x). \eeqn The axial charge $\tilde{Q}^A = \int dx \ \tilde{\rho}^A(x) $ is formally conserved in the field theory. Let us again consider the following partial axial rotation constructed based on the modified axial charge $\tilde{Q}^A$: \beqn && \exp\left( \ii \int_{-\infty}^x dx \frac{\pi}{N} \tilde{\rho}^A(x) \right) \cr\cr &=& \exp\left( \ii \int_{-\infty}^x dx \ \frac{\pi}{N} \rho^A(x) \right) \cr\cr &\times& \exp\left( \ii \int_{-\infty}^x dx \frac{e}{N} A(x) \right). \label{frac} \eeqn This operator is a product between the previous gauge invariant axial rotation at angle $\alpha = 2\pi/N$, with an extra fractional Wilson-line, which obviously violates the gauge invariance. There are various candidate operators on the lattice that reduce to $\tilde{Q}^A$ at long scale, but we will skip this discussion as $\tilde{Q}^A$ is not a valid operator to consider.

To restore the gauge invariance, we will need to combine the fractional Wilson-line in the last line of Eq.~\ref{frac} with another creation operator of a fractional gauge charge $e^\ast = -1/N$. Namely the last line of Eq.~\ref{frac} should be modified as \beqn \Psi^\dagger_{e^\ast}(x) \exp\left( \ii \int_{-\infty}^x dx \ \frac{e}{N} A (x) \right), \label{frac1} \eeqn where $\Psi^\dagger_{e^\ast}(x)$ is the creation operator of a state with gauge charge $e^\ast = - e / N$. $\Psi^\dagger_{e^\ast}(x)$ should come from the gapped ``dark sector" that also couples with $A(x)$. In the next subsection we will see that the holes in the system can play the role as the dark fermion $\psi_{\rm d}$, and provides a reservoir of fractional polarization charges.

\subsection{Axial charge $\cQ^\cA$ after restoring conservation and gauge invariance}

We have seen that the original axial charge $Q^A$ is gauge invariant but not conserved, while $\tilde{Q}^A$ is conserved but does not generate a gauge invariant axial rotation.

$ $

{\it Is there a ``proper" axial charge that is both conserved and also generates gauge invariant axial rotation?}

$ $

Let us not forget that we have so far only considered physics of electrons, which is only part of the system. Based on our review in the introduction, the total mechanical momentum of electrons and dark fermions $\psi_{\rm d}$ is $P^T = P + P_{\rm d}$, and it should be both conserved and gauge invariant, provided the system has translation invariance. We assume that $\psi_{\rm d}$ carries charge $ - e$, with the same filling factor $\nu = 1/N$. We can therefore define a proper axial charge $\cQ^\cA$ as \beqn P^T = \int dx \ \cP^T(x) = \frac{\pi}{N} \cQ^\cA = \int dx \ \frac{\pi}{N} \varrho^\cA(x). \eeqn We can also investigate the role of the following partial translation constructed with $P^T$: \beqn && \exp\left( \ii \int^x_{-\infty} dx \ \cP^T(x) \right) \cr\cr &=& \exp\left( \ii \int^x_{-\infty} dx \ \frac{\pi}{N} \rho^A (x) \right) \cr\cr &\times& \exp\left( \ii \int^x_{-\infty} dx \ \cP_{\rm d}(x) \right). \eeqn The last term in the equation above does take the desired form of Eq.~\ref{frac1} \beqn && \exp\left(\ii \int^x_{-\infty} dx \ \cP_{\rm d}(x) \right) \cr\cr &=& \exp\left( \frac{\ii}{N} \theta_{\rm d} (x) \right) \exp \left( \ii \int^x_{-\infty} dx \ \frac{e}{N} A(x) \right) \eeqn Compared with Eq.~\ref{frac}, this operator included another creation operator of fractional charge $\Psi^\dagger_{e^\ast}$ discussed in Eq.~\ref{frac1}: \beqn \Psi^\dagger_{e^\ast}(x) \sim \exp\left( \frac{\ii}{N} \theta_{\rm d} (x) \right)\eeqn which combined with the fractional Wilson-line in Eq.~\ref{frac} becomes gauge invariant. $\Psi^\dagger_{e^\ast}$ simply creates a fractional polarization charge $e^\ast = - e / N$, again through translation.

\subsection{Discussion}

Let us recap what we did. We first view the electrons as our ``target" physical system. Apparently the gauge invariant mechanical momentum, which is associated with the original axial charge, is not conserved, due to the Schwinger-anomaly, or simply the Newton's law. We then introduce the modified axial charge $\tilde{\rho}^A$ after naively restoring the conservation. Eq.~\ref{frac} is the axial rotation at angle $\alpha = 2\pi/N$ generated with the conserved axial charge $\tilde{\rho}^A$, but it cannot be made gauge invariant unless it is bound with the creation operator of another fractional gauge charge. In the model above we used holes as the dark fermion $\psi_{\rm d}$, which provides the ``reservoir" of the fractional charges, based on the physics that $1d$ systems at fractional filling supports fractional polarization charges under translation.

We have not made any assumptions about the dynamics of $\psi_{\rm d}$. If $\psi_{\rm d}$ is gapless, they cannot be ignored in our discussion of low energy modes of electrons, and the contribution of $\psi_{\rm d}$ to low energy physics should be taken into account from the very beginning. We should consider the scenario where $\psi_{\rm d}$ is gapped, then only electrons contribute to the low energy physics. However, since this is a $1d$ system, if $\psi_{\rm d}$ is gapped, it needs to spontaneously break the translation symmetry based on the Lieb-Shultz-Mattis theorem. If we want $\psi_{\rm d}$ at fractional filling to be gapped while preserving the translation symmetry, $\psi_{\rm d}$ needs to form a noninvertible topological order (a real dark sector), which cannot exist in $1d$ systems. In the next section we will discuss the mechanism of ``symmetry from the ABJ-anomaly" in a $3d$ model, there the dark sector can indeed be a noninvertible topological order.

Alternatively, one can consider the situation where a $1d$ conducting wire of electrons is coupled with a $2d$ system with $\psi_{\rm d}$. Then the $2d$ system can form a topological order, and resume the role of reservoir of fractional charges, without spontaneously breaking the translation. One natural scenario is for the $1d$ system to be the topologically protected boundary of a $2d$ bulk, which we will leave to a future work.

\section{A 3d model}

\subsection{Weyl semimetal and the ABJ-anomaly}

Let us start with a Bloch Hamiltonian of a Weyl-semimetal: \beqn && H(\vect{k}) = v \sin(k_x) \sigma^x + v \sin(k_y) \sigma^y \cr\cr && + v_z \left(\gamma - \cos(k_x) - \cos(k_y) - \cos(k_z) \right) \sigma^z. \label{weyl}\eeqn There are two Weyl fermions (which together form a Dirac fermion) at momentum $\vect{K}_{L,R} = (0, 0, \pm k_F)$, with $\cos(k_F) = \gamma - 2$. We assume that $k_F$ is small compared with the Brillouine zone boundary, therefore we can work in the continuum: \beqn H = \sum_{\vect{k}} \psi^\dagger_{\vect{k}} \left( v k_x \sigma^x + v k_y \sigma^y + \left(\frac{k_z^2}{2m} - \mu\right) \sigma^z \right) \psi_{\vect{k}}, \label{weyl}\eeqn and in this continuum model $k_F = \sqrt{2\mu m}$. We will still assume $k_F = \pi/N$ with integer $N$. In the Weyl semimetal, the axial rotation symmetry is the translation along $\hat{z}$, and as long as the charge U(1) symmetry and the translation along $\hat{z}$ ($T_z$) are preserved, the system remains gapless. When $k_F = \pi/N$, $T_z$ acts on the low energy modes as if it is a $Z_N$ axial symmetry. In particular, under translation $T_z$ by one lattice constant, the right and left fermions each acquire phase $\pm \pi/N$, and the fermion mass operator undergoes an axial rotation with $\alpha = 2\pi/N$.

\begin{figure}
    \centering
    \includegraphics[width=0.8\linewidth]{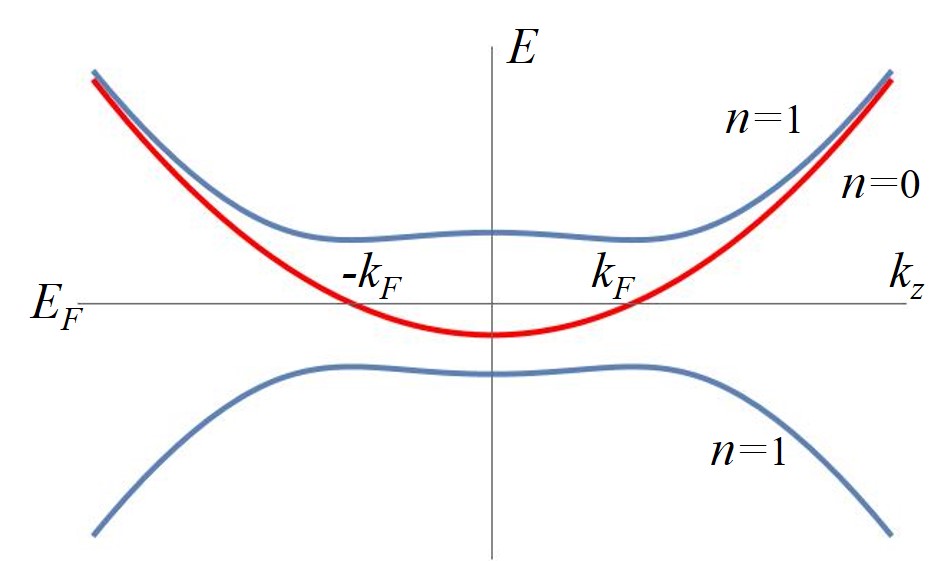}
    \caption{The energy bands of a Weyl semimetal in a uniform magnetic field along $\hat{z}$. The 0th Landau level $n = 0$ is the active band that contributes to the low energy physics, and it leads to $1d$ conducting channels which map to the $1d$ model in the previous section. }
    \label{dispersion}
\end{figure}

To get a physical picture of the ABJ-anomaly, we make one further simplification and consider electric and magnetic field in the $\pm \hat{z}$ direction, and the ABJ-anomaly should already be nontrivial in this case. Physically this is assuming the system has a very anisotropic dielectric function and magnetic permeability in space. We first consider the system in a uniform magnetic field along the $- \hat{z}$ direction. The system has a good translation symmetry along $ \hat{z}$, though in the XY plane it should have a more complicated magnetic translation. The Hamiltonian in the continuum can be solved (similar to a practice problem in Ref.~\onlinecite{peskin}): \beqn E_n = \pm \sqrt{\left( \frac{k_z^2}{2m} - \mu \right)^2 + 2 |B| v^2 e^2 n}, \ \ n = 1, 2, \cdots \eeqn The states with $|n| \geq 1$ form completely filled or empty bands, hence they should be viewed as the background insulator (Fig.~\ref{dispersion}). We can use the form of the Bloch wave function in the states with $|n| \geq 1$ to evaluate the charge polarization, and show that these filled bands do not give nontrivial response to electric field. Therefore we just need to investigate the 0th Landau-level $n = 0$ states. The $n = 0$ band has a definite chirality $\sigma^z = + 1$, and it connects the two $\vect{K}$ valleys, one valley with $E = v k_z$, the other valley with $E = - v k_z$.

Assuming the system has size $L\times L \times L_z$, there are in total $L^2 Be/(2\pi)$ channels of $n = 0$ modes connecting the two valleys, each channel is a $1d$ system at filling $\nu = k_F/\pi$. Now each channel at $n = 0$ precisely reduces to the previous model of $1d$ conductor. The total mechanical momentum is the sum of all channels at $n = 0$, and we can still project it at the Fermi wave vectors along this channel. Note that we only need to keep the $\sigma^z = + 1$ solutions. Then the evolution of $P_z$ is \beqn \frac{dP_z}{dt} = \sum_{c} \int dz  - e \frac{A_z(\vect{x})}{dt} \rho\cdots. \eeqn Here $\sum_c$ represents the sum over channels. each channel hosts fermion density $\nu = 1/N$ along the $\hat{z}$ direction. Again, $k_F = \pi \nu = \pi/N$.

The density of channels in the XY plane is $ \rho_{ch} = |B|e/(2\pi)$, hence the $3d$ density of electrons in the 0th Landau level is \beqn \rho = \nu \rho_{ch} = \nu |B|e/(2\pi). \eeqn Eventually the mechanical momentum of electrons becomes \beqn P_z = \int d^3x \ \frac{\pi}{N} (\rho_{R} - \rho_{L}) + \frac{e^2}{N}\frac{A_z(\vect{x}) B_z}{2\pi}. \label{Pz} \eeqn We can now define an axial charge $Q^A$ \beqn P_z = \frac{\pi}{N} Q^A = \int d^3x \frac{\pi}{N} \rho^A(\vect{x}). \cr\cr \rho^A = (\rho_{R} - \rho_{L}) + \frac{e^2}{2\pi^2} A_z B_z. \label{QA} \eeqn If the system is also in an electric field along $\hat{z}$,  the time-evolution of $P_z$ leads to the evolution of $Q^A$: \beqn \frac{dQ^A}{dt} = - \int d^3x \ \frac{e^2}{2\pi^2} E_z B_z. \eeqn This is the ABJ-anomaly equation. Hence the axial charge $Q^A$ (and the mechanical momentum $P_z$) is gauge invariant but not conserved. These results should still hold when $B_z$ weakly modulates in the XY plane, as the density of channels in the 0th Landau level is bound with $B_z(\vect{x})$.

One can define a modified axial charge which is naively conserved, by absorbing the ABJ-anomaly into the definition of the axial charge: $ \tilde{Q}^A = Q^A - \int d^3x \ \frac{e^2}{2\pi^2}A_z B_z$. But then the same problem arises: any axial rotation at angle $\alpha = 2\pi/N$ generated by $\tilde{Q}^A$ will violate the gauge invariance.

In order to define a conserved axial charge which also generates gauge invariant axial rotation at least with specific angles $\alpha = 2\pi p /N$, we again need to consider the total mechanical momentum \beqn P^T_z = P_z + P_{{\rm d},z}, \eeqn where $P_{\rm d,z}$ is the mechanical momentum of dark fermion $\psi_{\rm d}$ with charge $-e$. We first consider a Weyl semimetal band structure for $\psi_{\rm d}$, but with an opposite chirality from the electrons: \beqn && H_{\rm d}(\vect{k}) = v \sin(k_x) \sigma^x - v \sin(k_y) \sigma^y \cr\cr &+& v_z \left(\gamma - \cos(k_x) - \cos(k_y) - \cos(k_z) \right) \sigma^z. \label{weylhole}\eeqn In a magnetic field along $\hat{z}$, the electron and $\psi_{\rm d}$ both have 0th Landau level solutions with $n = 0$, and they have equal density and $k_F$ in the $n = 0$ channels. Hence within each channel, the system reduces to precisely the same problem considered in the previous section, with compensated electron and $\psi_{\rm d}$.

\begin{figure}
    \centering
\includegraphics[width=\linewidth]{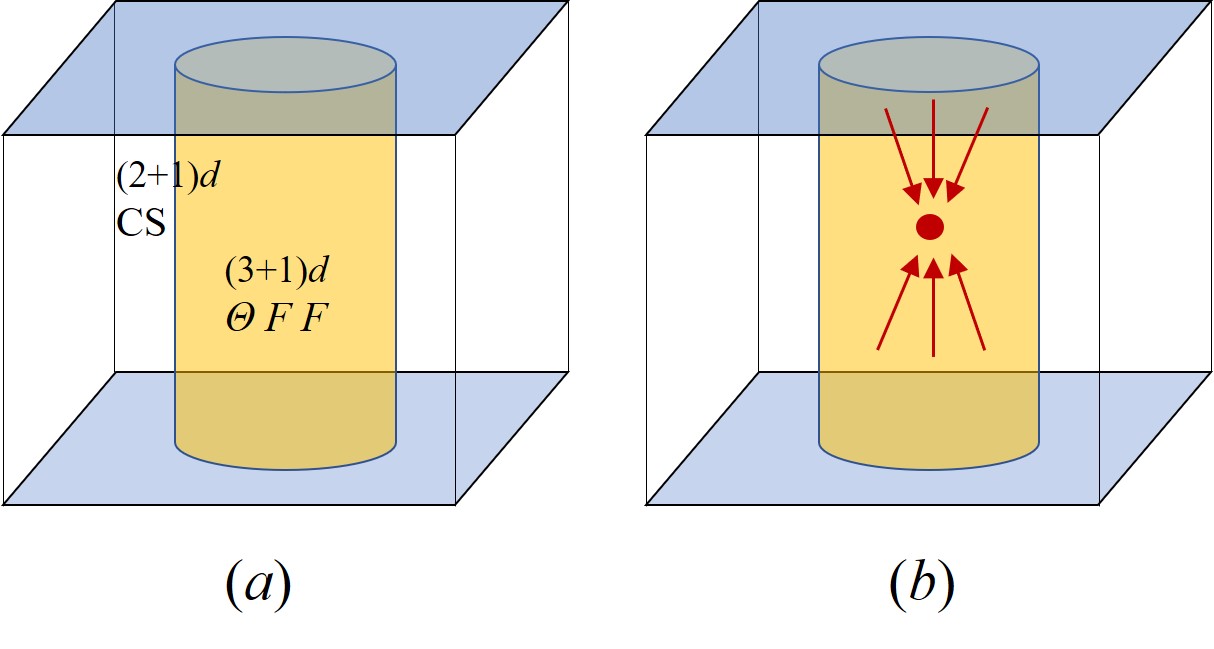}
    \caption{$(a)$ When performing partial translation in a cylindrical subregion $\cV$, the mechanical momentum of $\psi_{\rm d}$ generates a $\Theta$-term in $\cV$, which reduces to a CS term at the boundary $\partial \cV$, like Eq.~\ref{axialcs0}. But the total $\Theta$-term generated by translation from the entire electron+$\psi_{\rm d}$ system will cancel out. $(b)$ Similar effects occur to the charge carried by a magnetic monopole. When we perform translation on a monopole, the gapped dark sector will dress the monopole with charge $e^\ast = - e/N$, which cancels the contribution from the gapless QED of the electron. }
    \label{cylinder}
\end{figure}

The corresponding ``proper" axial charge can be defined as \beqn \cQ^\cA = Q^A + Q^A_{\rm d}. \eeqn 
The $\psi_{\rm d}$ contribution to the axial charge is $\frac{\pi}{N}Q^A_{\rm d} = P_{{\rm d},z}$, and it obeys the equation \beqn \frac{dQ^{A}_{\rm d}}{dt} = + \int d^3x
\frac{e^2}{2 \pi^2} \vect{E} \cdot \vect{B}. \eeqn 
The axial rotation at angle $\alpha = 2\pi/N$ generated by the proper axial charge $\pi\cQ^\cA/N$ is gauge invariant, as
$\pi\cQ^\cA/N$ is the total mechanical momentum of all the charged particles. If we perform a rotation with angle $\alpha = 2\pi/N$ in a subregion $\cV$ defined with $\cQ^\cA$, the mechanical momentum of dark fermion $P_{{\rm d},z}$ will generate a $\Theta$-term of the EM field in $\cV$, which precisely plays the same role as Eq.~\ref{axialcs0} after reducing to the boundary $\partial \cV$:
\beqn  && \exp\left( \frac{\ii }{2} \frac{2\pi}{N}  \int dt \int_{\cV} d^3x \ (\partial_\mu j^{\mu,A}_{\rm d}) \right) \cr\cr &=& \exp\left( \ii \int dt \int_\cV d^3x \ \frac{2\pi}{N} \frac{e^2}{32\pi^2} \epsilon^{\mu\nu\rho\sigma}F_{\mu\nu}F_{\rho\sigma} \right) \cr\cr &=& \exp\left( \ii \int  dt \int_{\partial \cV} d^2x \ \frac{e^2}{N} \frac{1}{4\pi} \epsilon^{\mu\nu\rho}A_{\mu} \partial_\rho A_{\sigma} \right). \label{axialcs2} \eeqn The reduction to the boundary CS term requires the assumption that the magnetic monopoles of $A_\mu$ are massive and invisible at low energy. The boundary CS term, though has a fractional level, should be gauge invariant, as it comes from a $\Theta$-term in the bulk of $\cV$ (Fig.~\ref{cylinder}).

\subsection{A $3d$ topological order}

So far the dark fermions $\psi_{\rm d}$ in the system are gapless, and we cannot exclude it from the discussion of low energy physics. If we want to ensure that the QED with one Dirac fermion of electrons is the low energy physics, we need to eliminate the gapless modes of $\psi_{\rm d}$. As we explained, the gapless Weyl fermions of a Weyl semimetal are protected by the translation symmetry $T_z$. When the system develops the following commensurate density wave along $\hat{z}$ with wave vector $2k_F$, $T_z$ is spontaneously broken, and the Weyl fermions are gapped: \beqn \langle \psi^\dagger \sigma^z \psi \rangle \sim \cos\left( 2 k_F (z - j) \right) \label{CDW}\eeqn Here $j = 0, 1, \cdots N - 1$. The density wave order parameter couples to the Weyl fermions as a mass term \beqn m \bar{\psi} e^{\ii \frac{2 \pi j}{N} \gamma_5} \psi. \label{axialmass}\eeqn

The density wave is the simplest way of gapping a Weyl semimetal. But this is certainly not ideal, as a density wave order parameter would couple to and gap out both the electron and $\psi_{\rm d}$ semimetals simultaneously, and alter the low energy dynamics of the $(3+1)d$ QED.

$ $

{\it We would like to construct a gapped state of $\psi_{\rm d}$ without breaking any translation symmetry, hence it does not gap out the electrons; but the gapped state of $\psi_{\rm d}$ should still preserve the desired $\Theta$-term in Eq.~\ref{axialcs2}}.

$ $

A natural state that meets these criteria is a topological order. Unlike the $1d$ model considered before, here $\psi_{\rm d}$ can indeed form a $3d$ topological order. Let us construct a specific topological order. {\it We eventually will need $N$-flavors of $\psi_{\rm d}$, but let us discuss one flavor first}. We will adopt the parton construction of fractionalized states analogous to Ref.~\onlinecite{fracTI1,fracTI2}, and introduce $N$ species of fermionic partons $\chi_a$ for each flavor of $\psi_{\rm d}$. Note that here $a$ is the color rather than flavor index. Each color of $\chi_a$ carries charge $ e^\ast = - e/N$ under the EM field $\vect{A}$. We also couple $\chi_a$ to a $\mathbb{Z}^A_N \times \mathbb{Z}^B_N$ dynamical gauge field. The $\chi_a$ carries charge $+1$ under the $\mathbb{Z}_N^A$ gauge field, and the $\mathbb{Z}_N^B$ permutes among different color species of partons. Note that in principle this construction allows $\chi_a$ couple to a $\SU(N)$ gauge field, but $\mathbb{Z}^A_N \times \mathbb{Z}^B_N $ is sufficient for our purpose. The gauge invariant bound state of $\chi_a$ is identified with $\psi_{\rm d}$: \beqn \psi_{\rm d} \sim \epsilon_{a_1 \cdots a_N} \chi_{a_1} \cdots \chi_{a_N}. \label{parton} \eeqn

We would like to stress that state of $\psi_{\rm d}$ itself is NOT close to a Weyl semimetal, and Eq.~\ref{parton} should be understood on the lattice rather than in the continuum. If one starts with a Weyl semimetal state of $\psi_{\rm d}$ and explore its proximate phases, the possible topological orders are much more restrictive, examples of such states are studied in Ref.~\onlinecite{wangburkov,chengweyl}. In fact, the theorem in Ref.~\onlinecite{cordovatheorem} precluded the emergence of a symmetric topological order from a Weyl semimetal except for special cases.

We begin with a mean field state of $\chi_a$: $\chi_a$ form a Weyl semimetal given by Eq.~\ref{weylhole} with a density wave along the $\hat{z}$ direction, with $j = a$ in Eq.~\ref{CDW} and Eq.~\ref{axialmass}. Though each color of fermion $\chi_a$ forms a density wave that breaks the translation, the entire system is still invariant under translation multiplied with a cyclic permutation of $\mathbb{Z}^B_N$ among the species. Hence the translation $T_z$ is still a symmetry of the system, realized as a projective symmetry group. With this mean field state, the $\mathbb{Z}_N^B$ gauge field is Higged, but the $\mathbb{Z}_N^A$ gauge field is still dynamical, and there is a $\mathbb{Z}_N^A$ gauge field for each flavor of $\psi_{\rm d}$.

Within the topological order of $\psi_{\rm d}$ discussed above, we consider action of the operator Eq.~\ref{axialcs2} in a cylindrical subregion $\cV$. For each parton color species, we can evaluate the ABJ-anomaly equation for its axial current: \beqn\partial_\mu j^{\mu,A}_a &=& \frac{1}{N^2} \times \frac{e^2}{16 \pi^2} \epsilon^{\mu\nu\rho\tau} F_{\mu\nu} F_{\rho\tau} \cr\cr &+& 2 m \ii \bar{\chi}_a \gamma_5 e^{\ii \frac{2 \pi a}{ N} \gamma_5} \chi_a. \label{partonj2} \eeqn The factor $1/N^2$ comes from the fact that each parton carries charge $e^\ast = - e/N$. The extra mass term in the second line comes from the mean field state of $\chi_a$, which has a commensurate density wave. The divergence of the total axial current is the sum of the axial current of partons within each flavor, then project to the $\mathbb{Z}_N^A \times \mathbb{Z}_N^B$ gauge invariant sector, and eventaully multiplied by the total flavor number $N$: \beqn \partial_\mu j^{\mu,A}_{\rm d} = N \times \hat{S} \left( \sum_{a = 1}^N \partial_\mu j^{\mu,A}_a \right) \hat{S}. \eeqn $\hat{S}$ performs the projection to the gauge invariant sector. The key observation is that, the last term of Eq.~\ref{partonj2} is not gauge invariant under the $\mathbb{Z}_N^A \times \mathbb{Z}_N^B$ gauge field. Therefore after projection, only the first line of Eq.~\ref{partonj2} remains. Then the result of Eq.~\ref{axialcs2} is recovered, even though $\psi_{\rm d}$ is in a gapped topological order. Since there is a $\mathbb{Z}_N$ topological order for each flavor of $\psi_{\rm d}$, the entire dark sector has a $(\mathbb{Z}_N)^N$ topological order.

The discussion above applies to odd integer $N$, as the dark particle is a fermion. But the disucssion can be straightforwadly generalized to the situation with even $N$, as long as we introduce a bosonic dark sector. We note that, Ref.~\onlinecite{shao2022} also proposed introducing an extra 2-form $\mathbb{Z}^{(2)}_N$ gauge field coupled to $\vect{A}$ in the $(3+1)d$ bulk with a twist as an alternative way of defining the proper axial symmetry.

\begin{figure}
    \centering
\includegraphics[width=0.9\linewidth]{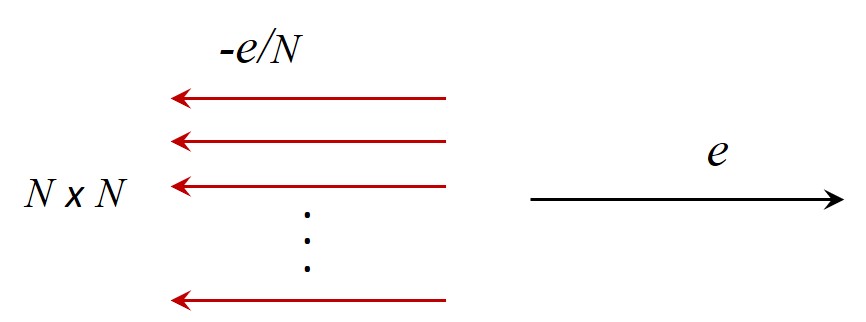}
    \caption{On the XZ or YZ boundary, the Weyl semimetal of electrons have one species of Fermi arc moving to the right. In the dark sector there are in total $N \times N$ species of partons Fermi arcs moving to the left, accounting for all the flavor and color species. }
    \label{arc}
\end{figure}

\subsection{Discussion and Physical Signatures}

Let us summarize our $3d$ model. Our model consists of electrons and dark fermion $\psi_{\rm d}$. We consider the situation where the electrons constitute the gapless Dirac fermion in the QED, while $\psi_{\rm d}$ forms a gapped $3d$ topological order. If we ignore $\psi_{\rm d}$, and only consider the axial charge of the gapless electrons, the axial rotation at any fractional angle $\alpha$ will not be gauge invariant. But if we consider the entire electron+$\psi_{\rm d}$ system, we can define a proper axial charge, which generates a gauge invariant axial rotation at angle $\alpha = 2\pi/N$, where $1/N$ is related to Weyl points as $k_F = \pi/N$.

The axial rotation of a naively defined ``conserved" axial charge $\tilde{Q}^A$ is not gauge invariant, and we need to introduce some extra degrees of freedom to fix the gauge transformation. The previous proposal was that~\cite{shao2022,cordova2022}, this can be fixed by introducing a FQH state on each domain wall of the action of axial rotation. Here we explored a simpler possibility: the extra degrees of freedom forms a topological order in the entire $3d$ system, rather than on each domain wall.

Let us now explore the physical signature of our electron+$\psi_{\rm d}$ system. The most prominent physical signature of a Weyl semimetal is the ``Fermi arcs", which leads to the quantum Hall effects. In our current set-up, the Fermi arcs exist at the XZ or YZ boundary. The electron and dark fermion $\psi_{\rm d}$ both have their own Fermi arcs at the boundary, and they have opposite chiralities. The mean field state of parton $\chi_a$ of $\psi_{\rm d}$ is designed in the way that there is no net charge Hall effect due to the Fermi arcs, as there are in total $N^2$ species of partons with charge $-e/N$ moving along the opposite direction from the electron.

We now consider the following magnetic monopole configuration of the $\vect{B}$ field: $2\pi/e$ flux quantum of the magnetic field along $\hat{z}$ terminates at coordinate $(0,0,z)$. The dark sector axial rotation Eq.~\ref{axialcs2} dresses the magnetic monopole with fractional polarization charge $e^\ast = - e/N$. This is still true in the topological order discussed above, after accounting for all the color and flavor species. Fractional polarization charge at a magnetic monopole occurs generally with a topological $\Theta$-term, as was discussed in previous works in the context of topological insulator~\cite{qi2008,moore2008}. Note that, the electrons and the dark sector will dress the monopole with opposite polarization charge, hence eventually the translation operator keeps the monopole charge neutral. All these are analogous to the heuristic example in section II.

In a Weyl semimetal, where are two physical consequences associated with the axial (translation) symmetry: the axial charge has the ABJ-anomaly, and a magnetic monopole collects nonzero polarization charge when translated along $\hat{z}$. But in the electron+$\psi_{\rm d}$ system we constructed, there is still an ABJ-anomaly for the gapless sector of the system, but the magnetic monopole no longer acquires net polarization charges, due to the cancellation from $\psi_{\rm d}$. However, we note that the gravitational anomaly of the entire electron+$\psi_{\rm d}$ system is not cancelled, which means that under a dynamical gravitational field, the conservation of the proper axial charge $\cQ^{\cA}$ will be broken. To further cancel the gravitational anomaly, another dark sector which is completely charge neutral needs be introduced.

\section{Outlook}

Translation is the most well-known realization of the axial symmetry in condensed matter systems, and in this work we explored how far one can extend the connection between translation and axial transformation, in the presence of gauge fields and anomaly. But translation is only similar to the axial symmetry at low energy, it is not an onsite $\U(1)$ symmetry in the microscopic model. In fact, the axial symmetry cannot be realized as an onsite $\U(1)$, unless the $(3+1)d$ QED is at the boundary of a $(4+1)d$ system. It is possible to explore a definition of ``proper" axial rotation with the presence of a $(4+1)d$ bulk. One possible direction is to construct a quantum spin Hall like state in the $(4+1)d$ bulk, where the spin $S^z$ is bound with the second Chern number of the gauge field $\vect{A}$, integrated over the $4d$ space of the bulk. One can potentially define an axial charge at the $(3+1)d$ boundary as the spin quantum number bound with the 2nd Chern number of gauge field in the bulk. The author leaves this direction to future exploration.

The Weyl semimetal as well as its proximate phases have many fascinating experimental signatures, such as the axion electrodynamics, which is similar to the pion electrodynamics~\cite{cordova2022}. These were explored in condensed matter literature in the past~\cite{weylburkov,weylphonon,weyltarun,weylson,weylson2}. In the future it will also be interesting to explore these experimental signatures in our electron+$\psi_{\rm d}$ state.

In our models the axial rotation is realized as the translation symmetry, which is an intrinsic symmetry of condensed matter systems, with or without interactions. We note that with a more specific Hamiltonian, other non-onsite symmetries can play the role as the axial symmetry, as was explored recently in Ref.~\onlinecite{gioia2025,shao2024}.

In the introduction we established several general criteria for the dark sector. The first criterion that demands the dark sector to be ``gapped without breaking any symmetry" strongly suggests a connection with another notion called ``{\it symmetric mass generation} (SMG)"~\cite{fidkowski1,Wang1307.7480,Wen_2013,SMG1,SMG2,Catterall1609.08541,ayyarMassive2015,Catterall1510.04153,ayyarOrigin2016,Tong2022SMG,he2016quantum,SMGdisorder,SMGAnna1,SMGAnna2,Simonanomaly,SMGreview}, which is precisely a mechanism of gapping out Dirac fermions without breaking the symmetry that precludes a simple Dirac mass term. We will also further explore this connection in the future.

The author thanks Meng Cheng and Shu-Heng Shao for very helpful discussion. We acknowledge support from the Simons foundation through the Simons investigator program. This research was inspired by the KITP program ``{\it Generalized Symmetries in Quantum Field Theory: High Energy Physics, Condensed Matter, and Quantum Gravity}", hence we also acknowledge supports by grant NSF PHY-2309135 to the Kavli Institute for Theoretical Physics (KITP). The author also acknowledges the use of generative AI, which saved time in calculations, and provided helpful references.

\appendix

\section{A $1d$ lattice model}

In the main text we mostly discussed Hamiltonians in the continuum. At least for $1d$, the Schwinger anomaly equation can be derived from a complete lattice model. Let us first consider a $1d$ tight-binding model of a charged fermion coupled with a $\U(1)$ gauge field: \beqn H &=& \sum_{j} - \frac{t}{2} ( e^{- \ii A_{j,j+1}} c^\dagger_{j+1} c_{j} + h.c.) + \frac{e E_{j,j+1}^2}{2}, \eeqn We define a ``dimensionless" current operator \beqn J &=& \sum_j - \frac{1}{t}\frac{\delta H}{\delta A_{j,j+1}} \cr\cr &=& \sum_{j} - \frac{1}{2} (\ii e^{- \ii A_{j,j+1}} c^\dagger_{j+1} c_{j}  + h.c.) \eeqn \beqn && \frac{dJ}{dt} = \ii [H, J] \cr\cr &=& \sum_j \frac{1}{2}  (e^{- \ii A_{j,j+1}} c^\dagger_{j+1} c_j + h.c. )  e E_{j,j+1}. \label{latticeJ} \eeqn Here $A_{j,j+1}$ and $E_{j,j+1}$ are quantum fields.

Now we project the equations above to the low energy modes of the system. Here the ``low energy" is not entirely well-defined, as a $1d$ lattie compact gauge field lead to confinement. We assume that $e$ is small enough, leading to a long confinement length, and our discussion applies to length scale no larger than the confinement length. To proceed, we make the following assumptions:

(1) We assume low filling factor, i.e. $\nu = 1/N \ll 1$, also
$k_F = \pi \nu \ll 1 $.

(2) We assume that the density of the system is uniform at low energy.

(3) We always expand to the leading nontrivial order of $A$ and $E$.

(4) We take $E$ and $A$ to be spatially uniform. The configuration of $E$ is bound with density of electric charges through the Gauss's law. Here we need to assume that the total electric charge density (including electrons and $\psi_{\rm d}$) is zero at low energy.

In this case, the current density is \beqn \frac{J}{L} &=&
\frac{1}{L}\sum_k \sin(k - A) c^\dagger_k c_k \cr\cr &\sim& k_F (\psi^\dagger_R \psi_R -\psi^\dagger_L \psi_L )- A \rho \cr\cr &\sim& \frac{\pi}{N} \rho^A, \eeqn where we have used the fact that $k_F = \pi/N$ and $\rho = 1/N$. Note that the definition of the gauge invariant $\rho^A$ includes gauge field $A$, which should be the case.

We also evaluate the last line of Eq.~\ref{latticeJ} at the Fermi wave vectors $\pm k_F$: \beqn && \sum_j \frac{1}{2} ( e^{- \ii A_{j,j+1}} c^\dagger_{j+1} c_j + h.c. ) \cr\cr &=& L \int \frac{dk}{2\pi} \cos(k - A) c_k^\dagger c_k \cr\cr &\sim& \frac{L}{\pi} \sin(k_F) \sim \frac{L}{N}. \eeqn Here we have used $c^\dagger_k c_k = 1 $ for $|k| < k_F$, and $c^\dagger_k c_k = 0 $ for $|k| > k_F$.

\beqn && \frac{d}{dt} \left( \frac{J}{L} \right) \sim \frac{1}{N} e E \cr\cr \ra \ \ \ &&  \frac{\pi}{N} \frac{d\rho^A}{dt} = \frac{1}{N} eE  \cr\cr \ra \ \ \  &&  \frac{d\rho^A}{dt} =
\frac{e}{\pi}E, \eeqn which is precisely the desired Schwinger anomaly equation.

In this $1d$ lattice model we have used the current operator rather than the momentum as the axial symmetry charge. When the filling is low, the current density operator will coincide with the momentum density. More generally, current density and momentum density are identical in systems with the Galilean symmetry.

\bibliography{big}

\end{document}